\documentclass[conference]{IEEEtran}

\usepackage{amsmath,amsfonts}
\usepackage{algorithmic}
\usepackage{algorithm}
\usepackage{array}
\usepackage{textcomp}
\usepackage{stfloats}
\usepackage{url}
\usepackage{verbatim}
\usepackage{graphicx}
\usepackage{cite}
\usepackage{xspace, subfigure, amssymb}
\usepackage[nolist,nohyperlinks]{acronym}
\usepackage{overpic}
\usepackage{lipsum}
\usepackage{xcolor}
\usepackage{multirow}

\newcommand{\sectionname}{Sect.}

\newcommand{\figuresname}{Figs.}

\newcommand{\meter}{\,\mathrm{m}}
\newcommand{\km}{\,\mathrm{km}}

\newcommand{\cm}{\,\mathrm{cm}}
\newcommand{\dB}{\,\mathrm{dB}}
\newcommand{\dBm}{\,\mathrm{dBm}}

\newcommand{\MHz}{\,\mathrm{MHz}}
\newcommand{\GHz}{\,\mathrm{GHz}}

\def\tr{\mathrm{tr}}

\def\Htran{\mathsf{H}}
\def\Ttran{\mathsf{T}}
\def\imagunit{\mathsf{j}} 
\newcommand{\vect}[1]{\boldsymbol{\mathbf{#1}}}

\newcommand{\Ex}[1]{\mathbb{E}\{{#1}\}}
\def\sinc{\mathrm{sinc}}

\begin{document}

\title{\acs{DFT}-Based Channel Estimation for Holographic~\acs{MIMO}}

\author{\IEEEauthorblockN{Antonio A. D'Amico, Giacomo Bacci, Luca Sanguinetti}
\IEEEauthorblockA{\textit{Dipartimento Ingegneria dell'Informazione, University of Pisa}, Pisa, Italy \\
\{antonio.damico, giacomo.bacci, luca.sanguinetti\}@unipi.it}\vspace{-0.5cm}
}

\maketitle

\begin{abstract}
  \Ac{hMIMO} systems with a massive number $N$ of individually controlled antennas make \ac{MMSE} channel estimation particularly challenging, due to its computational complexity that scales as $N^3$. This paper investigates uniform linear arrays and proposes a low-complexity method based on the \acl{DFT} approximation, which follows from replacing the covariance matrix by a suitable circulant matrix. Numerical results show that, already for arrays with moderate size (in the order of tens of wavelengths), it achieves the same performance of the optimal \ac{MMSE}, but with a significant lower computational load that scales as $N \log N$. Interestingly, the proposed method provides also increased robustness in case of imperfect knowledge of the covariance matrix.
\end{abstract}


\begin{acronym}[DVB-S2X]
\acro{BS}{base station}
\acro{CDF}{cumulative distribution function}
\acro{CSI}{channel state information}
\acro{DFT}{discrete Fourier transform}
\acro{EM}{electromagnetic}
\acro{KPI}{key performance indicator}
\acro{hMIMO}{holographic \acs{MIMO}}
\acro{mMIMO}{massive \acs{MIMO}}
\acro{i.i.d.}{independent and identically distributed}
\acro{ISO}{isotropic}
\acro{LoS}{line-of-sight}
\acro{MAI}{multiple access interference}
\acro{MIMO}{multiple-input multiple-output}
\acro{MMSE}{minimum mean square error}
\acro{NMSEE}{normalized mean square estimation error}
\acro{MR}{maximal ratio}
\acro{LoS}{line-of-sight}
\acro{LS}{least-squares}
\acro{NLoS}{non-line-of-sight}
\acro{RS-LS}{reduced-subspace \acs{LS}}
\acro{SE}{spectral efficiency}
\acro{SINR}{signal-to-interference-plus-noise ratio}
\acro{SNR}{signal-to-noise ratio}
\acro{UPA}{uniform planar array}
\acro{UL}{uplink}
\acro{ULA}{uniform linear array}
\acro{UE}{user equipment}
\acro{MIMO}{multiple-input multiple-output}
\acro{UaF}{use-and-then-forget}
\end{acronym}

\acresetall

\section{Introduction and motivation}\label{sec:intro}


Communication theorists are always on the lookout for new technologies to improve the speed and reliability of wireless communications. One such technology that has shown significant progress is multiple antenna technology, with the latest version being massive \ac{MIMO} \cite{massivemimobook}, which was introduced with the advent of 5G \cite{dsp_nextmMIMO}. Researchers are now exploring ways to deploy massive \ac{MIMO} with more antennas and optimized signal processing, given the potential benefits of numerous antennas. This technology evolution was named massive \ac{MIMO} 2.0 \cite{sanguinettiTCOM2020}, and new research directions are being pursued under different names, such as \ac{hMIMO} \cite{dsp_nextmMIMO,huang2020}, extremely large-scale \ac{MIMO} \cite{Niyato_2023}, and large intelligent surfaces \cite{Rusek_2018}.

The capacity of such technology evolution is theoretically unlimited \cite{sanguinettiTCOM2020}, but is practically limited, when the number of antennas increases, by the high computational complexity and the ability to learn the spatial channel correlation matrices.
%
In a \ac{hMIMO} system with thousands of antennas, it is challenging to both acquire the spatial correlation matrix and implement the \ac{MMSE} estimator \cite{demir2022WCL}.
The channel sparsity in the angular domain was exploited in \cite{wan2021} to perform channel estimation while reducing the pilot overhead, and \cite{cui2022} exploited the polar-domain sparsity when the angular-domain one is not applicable. To reduce the complexity, \cite{demir2022WCL} derives a subspace-based channel estimation approach based on the rank deficiency of the spatial correlation matrix caused by the \ac{hMIMO} geometry. In this case, the knowledge of the channel statistics is not required, and the complexity is reduced by considering isotropic scattering, which includes all possible angular spreads. 

Unlike the aforementioned literature, we propose a different low-complexity channel estimation scheme, based on the \ac{DFT}, and derived from a suitable circulant approximation of the channel covariance matrix \cite{Pearl1971,Pearl1973,Wakin2017}. Unlike \cite{demir2022WCL}, the estimation of the channel covariance matrix is required. To this aim, we also propose an improved, low-complexity algorithm to estimate the channel correlation matrix. Numerical results show that the proposed method provides almost the same accuracy of the optimal \ac{MMSE}
estimator, while significantly reducing the complexity thanks to the \ac{DFT} processing. This holds true for arrays of moderate size (order of tens of wavelenghts). Furthermore, when considering imperfect knowledge of the channel covariance matrix, the \ac{DFT}-based approach guarantees higher robustness and stability compared to the \ac{MMSE} method, thanks to a simpler eigenvalue structure. 


\section {System and channel model}\label{sec:system_model}

We consider a \ac{hMIMO} system where the \ac{BS} is equipped with a vertical \ac{ULA} located in the $yz$ plane, and consisting
of $N$ antennas, with inter-element spacing $d$~\cite[\figurename~1]{Sanguinetti2020}. The location of the $n$th antenna with respect to the origin is $\vect{u}_n = [ 0, 0, nd ]^{\Ttran}$, with $n=0,\dots,N-1$.\footnote{The analysis is valid for
any orientation of the \ac{ULA} with respect to the reference system, and can be extended to an horizontal \ac{ULA} straightforwardly.}
If a planar wave is impinging on the \ac{ULA} from the azimuth angle $\varphi$ and elevation angle $\theta$,
the array response vector is \cite[\sectionname~7.3]{massivemimobook}
\begin{align}\label{eq:array-response}
  \vect{a}(\varphi,\theta) = \left[e^{\imagunit\vect{k}(\varphi,\theta)^{\Ttran}\vect{u}_1},\dots,e^{\imagunit\vect{k}(\varphi,\theta)^{\Ttran}\vect{u}_N}\right]^{\Ttran}
\end{align}
where $ \vect{k}(\varphi, \theta) = \frac{2\pi}{\lambda}\left[\cos(\theta) \cos(\varphi), \, \cos(\theta) \sin(\varphi), \, \sin(\theta)\right]^{\Ttran}$ is the wave vector for a planar wave at wavelength $\lambda$. We call $\vect{h}_k\in \mathbb{C}^N$ the channel vector between the single-antenna \ac{UE} $k$ and the \ac{BS}, and assume that it consists of a superposition of multipath
components that can be expanded as a continuum of plane waves~\cite{Sayeed2002a}. Hence, we have
\begin{align} \label{eq:channel1}
  \vect{h}_k = \int_{-\pi/2}^{\pi/2} \int_{-\pi/2}^{\pi/2} g_k(\varphi,\theta) \vect{a}(\varphi,\theta) d\theta d\varphi 
\end{align}
where the \emph{angular spreading function} $g_k(\varphi,\theta)$ specifies gain and phase-shift from each direction
$(\varphi,\theta)$, and depends on \ac{UE} $k$'s position (i.e., angles $\varphi_k$ and $\theta_k$ with respect to the \ac{ULA}).

We consider the conventional block fading model, where the channel $\vect{h}_k$ is constant within
one time-frequency block and takes independent realizations across blocks from a stationary stochastic distribution.
In accordance with \cite{Sayeed2002a}, we model $g_k(\varphi,\theta)$ as a spatially uncorrelated circularly symmetric Gaussian
stochastic process with cross-correlation
\begin{align} \label{eq:scattering-correlation-model}
  \Ex{g_k(\varphi,\theta) g_k^*(\varphi^\prime,\theta^\prime)} = \beta_k f_k(\varphi,\theta) \delta(\varphi-\varphi^\prime)  \delta(\theta-\theta^\prime)
\end{align}
where $\beta_k$ is the average channel gain and $f_k(\varphi,\theta)$ is the normalized \emph{spatial scattering function} \cite{Sayeed2002a} such that $\iint f_k(\varphi,\theta) d\theta d\varphi  = 1$. By using \eqref{eq:array-response}, the
elements of $\vect{R}_k=\Ex{\vect{h}_k \vect{h}_k^{\Htran}}$ are computed as \cite[\sectionname~7.3.2]{massivemimobook}
\begin{align}
  [\vect{R}_k]_{m,l}  =  \beta_k \iint e^{\imagunit\vect{k}(\varphi,\theta)^{\Ttran} \left(\vect{u}_m - \vect{u}_l \right)} f_k(\varphi,\theta)d\varphi d\theta
\end{align}
where the integration is over all angles. If a vertical \ac{ULA} is used, the expression simplifies as
\begin{align}\label{eq:3D_local_correlation}
  [\vect{R}_{k}]_{m,l}  = \beta_k \int e^{\imagunit \frac{2\pi}{\lambda}d(m-l)\sin(\theta)} f_k(\theta)d\theta
\end{align}
where $f_k(\theta)=\int f_k(\varphi,\theta) d\varphi$. 

%

\section{Channel estimation with perfect knowledge}\label{sec:perfect_knowledge}


We assume that channel estimation is performed by using orthogonal pilot sequences of length $\tau_p$. We call 
$\vect{\phi}_{k} \in \mathbb{C}^{\tau_p}$ the pilot sequence used by \ac{UE} $k$ and assume that $|[\vect{\phi}_{k}]_i|^2 = 1$ and $\vect{\phi}_{k}^\Ttran \vect{\phi}_{k}^{*}  = \tau_{p}$.
In the absence of pilot contamination and with perfect knowledge of channel statistics, the linear \ac{MMSE} estimate of $\vect{h}_k$ based on the observation vector $\vect{y}_k=\tau_p \sqrt{\rho}\vect{h}_{k} + \vect{w}$, where $\rho$ is the transmit power, and $\vect{w} \sim \mathcal{CN}(\vect{0}, \tau_p\sigma^2 {\vect{I}}_N)$, is \cite[\sectionname~3]{massivemimobook}
\begin{align}\label{eq:mmse_estimator}
  \widehat{\vect{h}}_k^{\mathsf{MMSE}} = {\bf A}_k^{\mathsf{MMSE}}\vect{y}_k
\end{align}
where 
\begin{align}\label{eq:A-MMSE}
{\bf A}_k^{\mathsf{MMSE}}= \frac{1}{\tau_p \sqrt{\rho}}{\vect{R}}_k {\bf Q}_{k}^{-1} 
\end{align}
with ${\bf Q}_{k} = {\vect{R}}_k + \frac{1}{\gamma}{\vect{I}}_N$ 
and $\gamma = \tau_p\rho/\sigma^2$ denoting the \ac{SNR}.
The \ac{MMSE} channel estimator in \eqref{eq:mmse_estimator} is optimal, but it requires an intense
computational effort when $N$ grows large. Indeed, once ${\vect{R}}_k$ is computed, $\mathcal{O}(N^{3})$ operations are needed for the pre-computation of ${\bf A}_k^{\mathsf{MMSE}}$. The computation of \eqref{eq:mmse_estimator} requires a matrix-vector product evaluation whose complexity is $\mathcal{O}(N^{2})$. Its overall complexity is reported in Table \ref{tab:scheme}.
Note also that the \ac{MMSE} estimator relies on the perfect knowledge of ${\vect{R}}_k$, which needs to be estimated. 

An alternative estimation scheme is the \ac{LS}
estimator $\widehat{\vect{h}}_k^{\mathsf{LS}} = {\bf A}_k^{\mathsf{LS}}\vect{y}_k$, with
\begin{align}
\label{eq:A-LS}
{\bf A}_k^{\mathsf{LS}}= \frac{1}{\tau_p \sqrt{\rho}} {\bf I}_N 
\end{align}
which utilizes no prior information on the channel statistics and array geometry. Unlike the \ac{MMSE} channel estimator, it does not require any pre-computation phase and has a complexity of $\mathcal{O}(N)$, due to the product between the diagonal matrix ${\bf A}_k^{\mathsf{LS}}$ and $\vect{y}_k$. The price to pay is a reduced accuracy. 

\begin{table}[t]
\renewcommand{\arraystretch}{1.5}
\centering
\caption{Complexity of channel estimation schemes.}
\begin{tabular}{c|c|c}
\hline
{ \bf scheme} & {\bf pre-computation of} ${\bf A}_k$ & {\bf computation} of ${\bf A}_k{\bf y}_k$ \\
\hline
\acs{MMSE} & $\mathcal{O}(N^{3})$ & $\mathcal{O}(N^{2})$ \\
\acs{LS} & -- & $\mathcal{O}(N)$ \\
\acs{LoS} & -- & $\mathcal{O}(N)$ \\
\acs{ISO} & -- & $\mathcal{O}(N^{2})$ \\
\acs{DFT} & $\mathcal{O}(N \log N)$ & $\mathcal{O}(N \log N)$ \\
\hline
\end{tabular}
\label{tab:scheme}
\end{table}

Two other alternatives are described next and can be applied in specific propagation conditions. If propagation is assumed to take place in a \ac{LoS} scenario with a single plane-wave arriving from $\theta_k$ and $\varphi_k$, then $\vect{h}_k = g_k(\varphi_k,\theta_k) \vect{a}(\varphi_k,\theta_k) $ and ${\vect{R}}^{\mathsf{LoS}}_k=\beta_k\vect{a}(\varphi_k,\theta_k)\vect{a}(\varphi_k,\theta_k)^{\Htran}$.
Replacing $\vect{R}_k$ with ${\vect{R}}^{\mathsf{LoS}}_k$ into \eqref{eq:mmse_estimator} yields $\widehat{\vect{h}}_k^{\mathsf{LoS}} = {\bf A}_k^{\mathsf{LoS}}\vect{y}_k$, where
\begin{align}\label{eq:mmse_estimator_LoS}
  {\bf  A}_k^{\mathsf{LoS}} = \frac{1}{\tau_p \sqrt{\rho}}\frac{\beta_k \gamma}{1 +N\beta_k \gamma}\vect{a}(\varphi_k,\theta_k)\vect{a}(\varphi_k,\theta_k)^{\Htran}
\end{align}
whose complexity is $\mathcal{C}_{\mathsf{LoS}}=\mathcal{O}(N)$, due to the evaluation of the product between ${\bf  A}_k^{\mathsf{LoS}}$ in \eqref{eq:mmse_estimator_LoS} and ${\vect y}_k$ (no pre-computation phase is required). However, the \ac{LoS}-based estimator works well only when the channel vector is generated by a single plane-wave arriving from $(\varphi_k,\theta_k)$, whose knowledge must be perfectly known at the BS. When the propagation scenario is highly scattered, and plane waves arrive uniformly within the angular domain in front of the \ac{ULA}, we can make
use of the \ac{ISO} approximation proposed in \cite{demir2022}. According to \cite{demir2022}, ${\vect{R}}^{\mathsf{ISO}}_k= \overline{\vect{U}} \overline{\vect{\Lambda}} \overline{\vect{U}}^{\Htran}$
where $\overline{\vect{U}}$ and $\overline{\vect{\Lambda}}$ are the (reduced-order) eigenvector and eigenvalue matrices, obtained through
the \emph{compact} eigenvalue decomposition of a matrix whose $(m,l)$th entry is
$\sinc\left[2\left(m-l\right)d/\lambda\right]$, with $\sinc(x)=\sin(\pi x)/(\pi x)$. Note that the rank of ${\vect{R}}^{\mathsf{ISO}}_k$ is approximately $2Nd/\lambda$, given by the degrees of freedom
observed in the \ac{ISO} propagation conditions \cite{PizzoTSP22}. Replacing $\vect{R}_k$ with ${\vect{R}}_k^{\mathsf{ISO}}$ into \eqref{eq:mmse_estimator} yields $  \widehat{\vect{h}}_k^{\mathsf{ISO}} = {\bf A}_k^{\mathsf{ISO}}\vect{y}_k$,
with
\begin{align}\label{eq:12}
  {\bf A}_k^{\mathsf{ISO}} = \frac{1}{\tau_p \sqrt{\rho}} \overline{\vect{U}} \, \overline{\vect{\Lambda}} \left(\overline{\vect{\Lambda}} +\frac{1}{\gamma}{\vect{I}}_N \right)^{-1}\overline{\vect{U}}^{\Htran}.
\end{align}
Compared to \ac{MMSE}, the main advantage of using the \ac{ISO} estimator derives from the fact that it does not require any matrix estimation and inversion, since all the quantities in ${\bf A}_k^{\mathsf{ISO}}$ are known.
Accordingly, its complexity is only due to the matrix-vector product computation between ${\bf A}_k^{\mathsf{ISO}}$ and $\vect{y}_k$ and is $\mathcal{C}_{\mathsf{ISO}}=\mathcal{O}(N^2)$. Note also that the \ac{ISO} estimator does not require any prior knowledge of the channel statistics and can be applied to any propagation conditions, since the eigenspace of ${\vect{R}}_k^{\mathsf{ISO}}$ covers the eigenspace of any spatial correlation matrix ${\vect{R}}_k$ \cite{demir2022}, and exploits the array geometry only.

\subsection{\Acl{DFT} approximation}\label{perfect_knowledge:DFT}

We now develop a channel estimation scheme that exploits the correlation induced by the array geometry and propagation conditions to approach \ac{MMSE} performance, while having a computational complexity that scales log-linearly with $N$. To this aim, we proceed as follows.

\begin{figure}[t]
  \begin{center}\vspace{-0.5cm}
    {\includegraphics[width=\columnwidth]{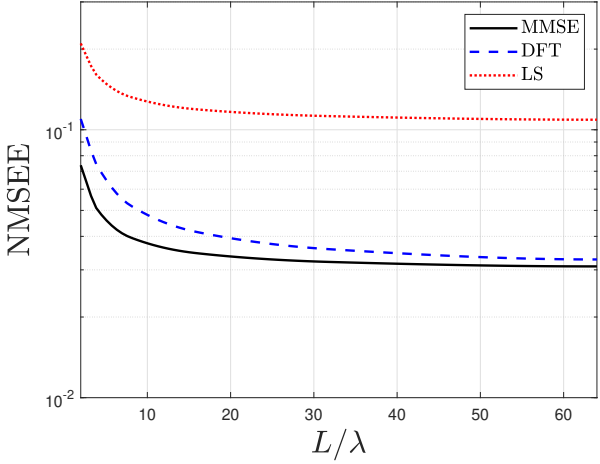}}
    \caption{\acs{NMSEE} as a function of the ratio $L/\lambda$.}
    \label{fig:perfect_NMSEE_vs_N}
  \end{center}
\end{figure}

If an \ac{ULA} is used, the covariance
matrix  $\vect{R}_k$ is Hermitian Toeplitz, and it can be approximated with a suitable circulant matrix
$\vect{C}_k$ \cite{Pearl1971,Pearl1973,Wakin2017}, whose first row $\vect{c}_{k}=[c_k(0),c_k(1),\cdots,c_k(N-1)]$ is related to the first
row $\vect{r}_{k}=[r_k(0),r_k(1),\cdots,r_k(N-1)]$ of $\vect{R}_k$ by \cite{Wakin2017}
\begin{align}\label{eq:Ck}
  c_k(n)=\begin{cases}
  r_k(0)& \text{$n=0$}, \\
  \dfrac{(N-n) r_k(n)+n r^{*}_k (N-n)}{N} & \text{$n = 1,\ldots,N-1$}.
  \end{cases}
\end{align}
Any circulant matrix can be unitarily diagonalized using the \ac{DFT} matrix, i.e., $ {\vect{C}}_k={\vect{F}} {\vect{\Lambda}}_{k}  {\vect{F}}^{\Htran}$
where ${\vect{F}}=[{\vect{f}}_{0} \, {\vect{f}}_{1} \, \cdots \, {\vect{f}}_{N-1}]$ is the inverse \ac{DFT} matrix, with
$[{\vect{f}}_n]_m=\left.e^{\imagunit 2 \pi mn/N}\right/\sqrt{N}$ for $0 \le m,n \le N-1$, and ${\vect{\Lambda}}_{k}$ is the diagonal matrix containing the eigenvalues of ${\vect{C}}_k$, i.e.,
\begin{align}
\label{DFT-firstrow}
  [{\vect{\Lambda}}_k]_{n,n}=\sum\limits_{m=0}^{N-1}c_{k}(m)e^{-\imagunit 2 \pi mn/N}  
\end{align}
which are obtained by taking the \ac{DFT} of the first row of $\vect{C}_k$. Replacing $\vect{R}_k$ with $\vect{C}_k$ into \eqref{eq:mmse_estimator} yields $\widehat{\vect{h}}_k^{\mathsf{DFT}} = {\bf A}_k^{\mathsf{DFT}}\vect{y}_k$,
with
\begin{align}
  {\bf A}_k^{\mathsf{DFT}} = \frac{1}{\tau_p \sqrt{\rho}} {\vect{F}} \, {\vect{\Lambda}_k} \left({\vect{\Lambda}_k} +\frac{1}{\gamma}{\vect{I}}_N \right)^{-1}{\vect{F}}^{\Htran}.
\end{align}
We call it the \emph{\ac{DFT}-based channel estimator}. Its complexity derives from the pre-computation phase, which is $\mathcal{O}(N \log N)$ due to the computation of ${\vect{\Lambda}}_k$ through \eqref{DFT-firstrow}, and from the computation of the matrix-vector product, which is again $\mathcal{O}(N \log N)$, since the \ac{DFT} matrix $\vect F$ and its inverse are involved. Hence, the  complexity of the \ac{DFT}-based estimator is $\mathcal{C}_{\mathsf{DFT}}=\mathcal{O}(N \log N)$. Unlike the \ac{ISO} estimator, the \ac{DFT}-based estimator depends on the true covariance matrix ${\vect R}_k$, which must be estimated as with the \ac{MMSE} estimator.

\begin{figure}[t]
  \begin{center}\vspace{-0.5cm}
    {\includegraphics[width=\columnwidth]{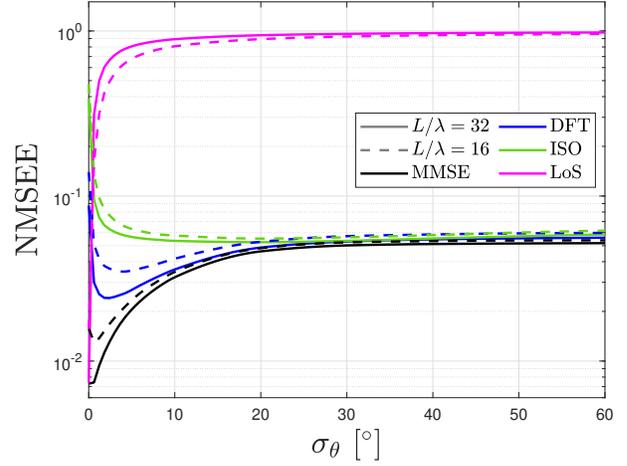}}
    \caption{\acs{NMSEE} as a function of $\sigma_\theta$.}
    \label{fig:perfect_NMSEE_vs_sigma}
  \end{center}
\end{figure}

\subsection{Performance analysis}\label{perfect_knowledge:performance}


\figurename~\ref{fig:perfect_NMSEE_vs_N} shows the \ac{NMSEE}, given by \cite[\sectionname~3]{massivemimobook} 
\begin{align}\label{eq:NMMSE}
\frac{\tr( \vect{R}_k ) - 2 \sqrt{\rho} \tau_p \Re\left( \tr( \vect{R}_{k} \vect{A}_k )  \right)
  + \rho\tau_p^2 \tr \left( \vect{A}_k \vect{Q}_{k} \vect{A}_k^{\Htran} \right)}{\tr( \vect{R}_k )}
\end{align}
where $\vect{A}_k$ depends on the estimation strategies defined above (e.g., \eqref{eq:A-MMSE} for \ac{MMSE}, and \eqref{eq:A-LS} for
\ac{LS}), as a function of the array size $L$, normalized with respect to the wavelength
$\lambda$. We consider an \ac{ULA} characterized by $d=\lambda/4$ at $3\GHz$ (and hence $\lambda=10\cm$). Three estimation schemes are considered:
\ac{MMSE}, \ac{LS}, and \ac{DFT}.
We evaluate the average performance for a \ac{UE} randomly placed in the sector $\varphi\in[-\pi/3, +\pi/3]$ of a cell with minimum and
maximum distances from the \ac{ULA} of $5$ and $100$ meters, respectively. The \ac{ULA} is elevated by $b=10\meter$ with respect to the \ac{UE}
plane (and thus, considering the distance range, $\theta\in[-63.4^\circ, -5.7^\circ]$, with negative elevations due to the fact that the \ac{UE} plane is below the \ac{ULA}),
and the received \ac{SNR} is $\beta_k\tau_p\rho/\sigma^2$, where: $\beta_k$ is computed following \cite[\sectionname~2]{massivemimobook}
assuming a reference distance of $1\km$, a path loss exponent $\alpha=3.76$, and a channel gain at $1\km$ equal to $-148.1\dB$;
$\tau_p=10$; $\rho=20\dBm$; and $\sigma^2=-87\dBm$, obtained considering a communication bandwidth $B=100\MHz$.
We assume a local scattering model with a Laplacian distribution characterized by angular scattering spread $\sigma_{\theta}=10^{\circ}$.
We see that the accuracy of the \ac{DFT}-based estimator is comparable with the optimal \ac{MMSE} one, and the gap decreases as $L/\lambda$ increases. This is due to the fact that the circulant approximation $\vect{C}_{k}$ of the covariance matrix $\vect{R}_{k}$
improves as the number of antennas $N$ (or, equivalently, the ratio $L/\lambda$) grows large.
Interestingly, the circulant approximation is already quite tight for $L/\lambda=16$ ($L=1.6\meter$ and $N=64$). More importantly, this is obtained with a complexity of $\mathcal{O}(N \log N)$, instead of $\mathcal{O}(N^3)$. If $N=64$, this corresponds to two orders of magnitude of computational saving 
compared to \ac{MMSE}.

To evaluate the impact of the angular spread, \figurename~\ref{fig:perfect_NMSEE_vs_sigma} plots the \ac{NMSEE} as a function
of $\sigma_{\theta}$ in the same simulation setup of
\figurename~\ref{fig:perfect_NMSEE_vs_N}. The results show that the \ac{DFT}-based estimator significantly outperforms (with a gap that increases with $L/\lambda$) both the \ac{LoS} and the \ac{ISO}-based estimators for values of $\sigma_{\theta}$ in the range $\left(5^\circ, 20^\circ\right)$ and attains good performance compared to the (optimal) \ac{MMSE}. As expected, the \ac{LoS} estimator is close to optimal only for very low values of $\sigma_{\theta}$. 

\section{Channel estimation with imperfect knowledge}\label{sec:imperfect_knowledge}

So far, we have assumed perfect knowledge of $\vect{R}_{k}$. This may not be the case in practical scenarios since $\vect{R}_{k}$ changes due to different reasons \cite{sanguinettiTCOM2020}. Measurements suggest roughly two orders of magnitude slower variations  compared to the
fast variations of channel vectors.
Therefore, we may reasonably assume that they do not change over $\tau_s$ coherence blocks, where $\tau_s$
can be in the order of thousands \cite{sanguinettiTCOM2020}. 
Suppose that the \ac{BS} has received the pilot matrix $\vect{y}_k$ in $M \le \tau_s$ coherence
blocks. We denote these $M$ observations by $\vect{y}_k[1],\ldots,\vect{y}_k[M]$. An estimate of ${\vect{Q}}_{k}$ can be obtained by computing the sample correlation matrix given by
\begin{align}\label{eq:Qsample}
  \widehat{\vect{Q}}_{k}^{\mathsf{sample}} = \frac{1}{M} \sum_{m=1}^{M} \vect{y}_k[m] \vect{y}_k^{\Htran}[m].
\end{align}
A better estimator is typically obtained through matrix regularization by computing the convex combination \cite{sanguinettiTCOM2020}:
\begin{align}\label{eq:Qestimate}
  \widehat{\vect{Q}}_{k}(\eta) = \eta\widehat{\vect{Q}}_{k}^{\mathsf{sample}}+ (1-\eta) \widehat{\vect{Q}}_{k}^{\mathsf{diag}} \quad  \eta \in [0,1]
\end{align}
where $\widehat{\vect{Q}}_{k}^{\mathsf{diag}}$ contains the main diagonal of $\widehat{\vect{Q}}_{k}^{\mathsf{sample}}$.
Once $\widehat{\vect{Q}}_{k}(\eta)$ is computed, an estimate of ${\vect{R}}_k$ follows:
\begin{align}
  \widehat{\vect{R}}_k(\eta) = \widehat{\vect{Q}}_{k}(\eta) - \frac{1}{\gamma}{\vect{I}}_N
\end{align}
which requires only knowledge of $\gamma$, i.e., the \ac{SNR} during the pilot transmission phase.

\subsection{Improved estimation of the channel correlation matrix}\label{imperfect_knowledge:improved}
We now develop an improved estimation scheme of the correlation matrix $\vect{Q}_k$ that can be used with
\acp{ULA}. In this case, $\vect{Q}_k$ is Hermitian Toeplitz, i.e.,
\begin{align}\label{eq:Q_toep}
  [\vect{Q}_k]_{1,j}=[\vect{Q}_k]_{1+m,j+m} 
\end{align} 
for $j=1,\ldots,N-1$ and $m=1,\ldots,N-j$, and $[{\vect{Q}}_k]_{i,j}=[{\vect{Q}}_k]^{\ast}_{j,i}$. To proceed, we denote
by $\widehat{\vect{Q}}^{\mathsf{toe}}_k$ the estimate of $\vect{Q}_k$ obtained by taking the Toeplitz structure \eqref{eq:Q_toep}
into account. The first row of $\widehat{\vect{Q}}^{\mathsf{toe}}_k$ is computed by simply averaging the entries of
$\widehat{\vect{Q}}_{k}^{\mathsf{sample}}$ in \eqref{eq:Qsample} over the diagonals, i.e.,
\begin{align}\label{eq:20}
[\widehat{\vect{Q}}^{\mathsf{toe}}_k]_{1,j}=\dfrac{1}{N-j+1} \sum\limits_{m=1}^{N-j+1}[\widehat{\vect{Q}}_{k}^{\mathsf{sample}}]_{m,j+m-1}.
\end{align} 
Once the first row is computed, the other elements are easily found. In particular, from \eqref{eq:Q_toep} we have that
\begin{align}\label{eq:Qs_toep}
  [\widehat{\vect{Q}}^{\mathsf{toe}}_k]_{1+m,j+m} =[\widehat{\vect{Q}}^{\mathsf{toe}}_k]_{1,j}
\end{align}   
for $j=1,\ldots,N-1$ and $m=1,\ldots,N-j$, and
\begin{align}\label{eq:Qs_herm}
  [\widehat{\vect{Q}}^{\mathsf{toe}}_k]_{j,i}=[\widehat{\vect{Q}}^{\mathsf{toe}}_k]^{\ast}_{i,j} \quad \quad \textrm{for $j >i$}
\end{align} 
because of the Hermitian symmetry of the covariance matrix. An estimate of $\vect{R}_k$ is finally obtained as
\begin{align}\label{eq:Rs_toep}
  \widehat{\vect{R}}^{\mathsf{toe}}_k=\widehat{\vect{Q}}^{\mathsf{toe}}_k-\frac{1}{\gamma}{\vect{I}}_N.
\end{align}
The complexity of the estimator above is mainly due to the computation of $\widehat{\vect{Q}}_{k}^{\mathsf{sample}}$ in \eqref{eq:20} and
thus is comparable to the one not exploiting the Toeplitz structure. We can then replace the elements of $\vect{R}_k$ with the ones in
$\widehat{\vect{R}}^{\mathsf{toe}}_k$ to implement \eqref{eq:Ck}, when considering the proposed \ac{DFT}-based estimator with imperfect
knowledge of channel statistics.

\begin{figure}[t]
  \begin{center}\vspace{-0.5cm}
    \subfigure[\acs{MMSE} estimator.]{
    {\includegraphics[width=\columnwidth]{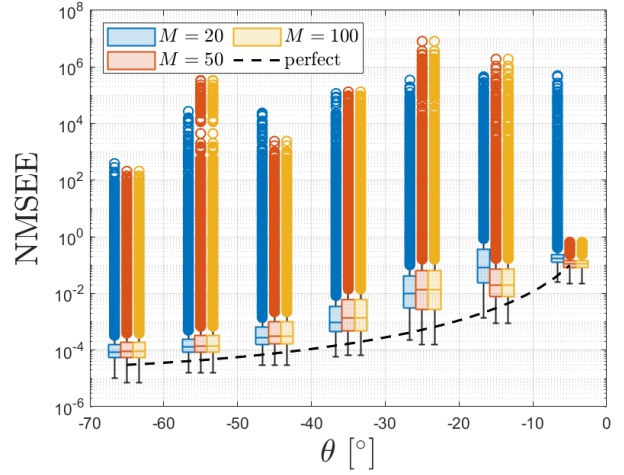}}
    \label{fig:imperfect_NMSEE_vs_elevation_MMSE}}
    \\
    \subfigure[\acs{DFT}-based estimator.]{
    {\includegraphics[width=\columnwidth]{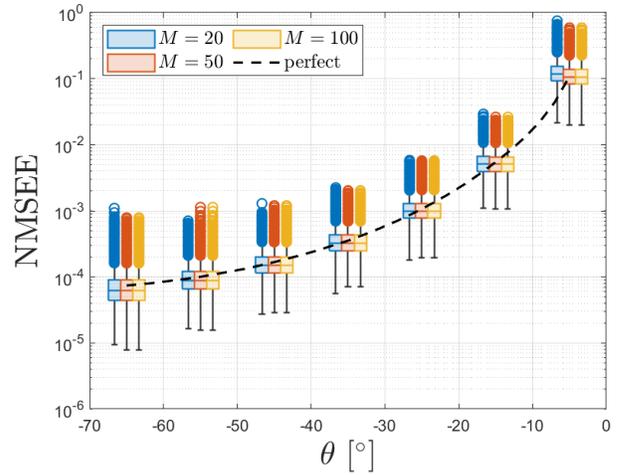}} 
    \label{fig:imperfect_NMSEE_vs_elevation_DFT}}
    \caption{Box plot of the \acs{NMSEE} as a function of the elevation $\theta$ with imperfect statistical knowledge ($N=64$).}
    \label{fig:imperfect_NMSEE_vs_elevation}
  \end{center}
\end{figure}

\begin{figure}[t]
  \begin{center}\vspace{-0.5cm}
    {\includegraphics[width=\columnwidth]{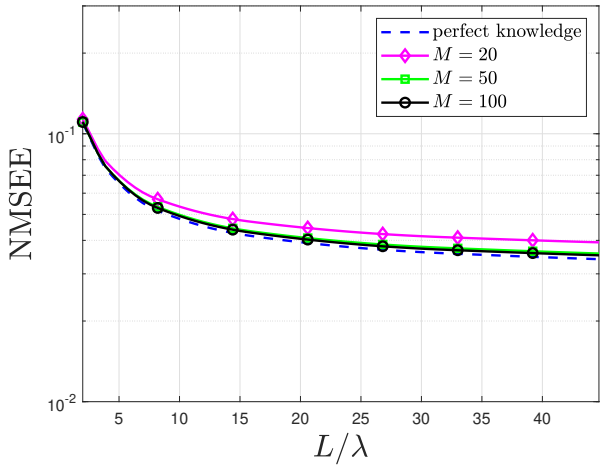}}
    \caption{\acs{NMSEE} as a function of the ratio $L/\lambda$ with imperfect statistical knowledge
      (\acs{DFT}-based estimator).}
    \label{fig:imperfect_NMSEE_vs_N_DFT}
  \end{center}
\end{figure}

\subsection{Performance analysis}\label{imperfect_knowledge:performance}
We now evaluate the accuracy of the estimators when the covariance matrix is estimated using \eqref{eq:Rs_toep}. 
\figuresname~\ref{fig:imperfect_NMSEE_vs_elevation}\subref{fig:imperfect_NMSEE_vs_elevation_MMSE} and
\subref{fig:imperfect_NMSEE_vs_elevation_DFT} report the box charts for the \ac{MMSE} and the \ac{DFT}-based estimators,
respectively, showing median, lower and upper quartiles, minimum and maximum non-outlier values, and outliers (the latter
depicted by circular markers), computed over $200,000$ independent realizations per box.
Blue, red, and yellow boxes correspond to the different values of $M$ considered for the
estimation of the covariance matrix: $M=20$, $50$ and $100$, respectively. The system setup is the same considered in
\sectionname~\ref{perfect_knowledge:performance}. 
For comparison, we also report the \ac{NMSEE} with
perfect knowledge of $\vect{R}_{k}$ (dashed line). By inspecting \figurename~\ref{fig:imperfect_NMSEE_vs_elevation}, the following
considerations can be drawn: the average estimation accuracy improves as the (absolute) elevation increases, thanks
to a reduced distance (which is related to the elevation angle through $b/|\sin \theta|$), and
thus to an increased \ac{SNR}. However, especially for the \ac{MMSE} case, the robustness of the estimator decreases
as the (absolute) elevation increases, owing to the reduced array directivity at large elevations.
This is confirmed by the huge presence of outliers, which highly affect the reliability of the \ac{MMSE} estimation, especially
when $M$ decreases. This result is somewhat expected, as we are using a reduced set of snapshots compared to the degrees of
freedom offered by the $N$-sized \ac{ULA}, which prevents from a stable and accurate estimation of the channel statistics.

Most interestingly, although the same trends apply to both \figuresname~\ref{fig:imperfect_NMSEE_vs_elevation}\subref{fig:imperfect_NMSEE_vs_elevation_MMSE} and
\subref{fig:imperfect_NMSEE_vs_elevation_DFT}, a significant difference can be observed when focusing on the \ac{DFT}-based estimator.
As can be seen, not only the average performance is close to the one with perfect knowledge, but also the standard deviation is orders of
magnitude lower than the \ac{MMSE} counterpart, and so does the number of outliers, already at $M=20$. This is due to a simpler estimation
scheme, which requires a reduced number of independent realizations, and thus exhibits larger robustness.
To provide further insights, \figurename~\ref{fig:imperfect_NMSEE_vs_N_DFT} reports the average \ac{NMSEE} (which also
includes the outliers) as a function of the ratio $L/\lambda$, obtained by averaging over all possible \ac{UE} positions in the range
$[5, 100]\meter$ and $\varphi\in[-\pi/3,+\pi/3]$, and using the same system parameters considered for \figurename~\ref{fig:imperfect_NMSEE_vs_elevation}.
As can be seen, an estimation accuracy comparable to that obtained with perfect knowledge of $\vect{R}_{k}$ is already achieved
with $M=20$.
Similar conclusions can be drawn by considering different simulation setups (not reported for space limitations), in which
different array sizes and/or scattering scenarios are considered.

\section{Conclusion}\label{sec:conclusion}

We proposed a low-complexity scheme, based on the circulant approximation of the channel covariance matrix, to perform channel estimation in \ac{hMIMO} systems equipped with \acp{ULA}. The estimation accuracy was evaluated with perfect and imperfect knowledge of the channel covariance matrix. Comparisons were made against the optimal \ac{MMSE} estimator and other alternatives with lower complexity. The proposed scheme achieves close to optimal estimation accuracy for \acp{ULA} of moderate size (in the order of tens of wavelength), while considerably reducing the estimation complexity by a factor that scales with the square of array size. Moreover, it is more robust to the imperfect knowledge of channel statistics. This makes it more suited for applications in which the statistics changes rapidly over time and must be estimated frequently using a limited number of coherence blocks. Future work is needed to extend the proposed scheme to uniform planar arrays.

\bibliographystyle{IEEEtran}
\bibliography{mybib}

\end{document}